\documentstyle[fleqn,12pt]{article}
\author{L.~Didukh and V.~Hankevych \\
{\small \it Ternopil State Technical University, Department of Physics,}\\ 
{\small \it 56 Rus'ka Str., Ternopil UA--282001, Ukraine; Tel.:+380352251946,}
\\ 
{\small \it Fax: +380352254983; E-mail: didukh@tu.edu.te.ua}}

\title{Metal-insulator transition in a generalized Hubbard model with 
correlated hopping at half-filling}
\date{}
		
\begin{document}

\maketitle
\begin{abstract}
In the present paper metal-insulator transition is studied in a generalized
Hubbard model with correlated hopping at half-filling and zero temperature.
Single-particle Green function and energy spectrum of electron system are
calculated. The expressions for energy gap width and the concentration of
polar states (holes or doublons) are obtained. The conditions for metallic
and insulating states are found.

PACS number(s): 71.30.+h, 71.10.Fd, 71.27.+a, 71.28.+d, 
\end{abstract}

\section{Introduction}

In the present paper we consider a generalized Hubbard model including 
correlated hopping~\cite{1,2}. In such a model the hopping integrals, 
which describe hoppings of
holes and doublons are different. These hopping integrals also are 
distinguished
from the hopping integral which is connected with the processes of paired 
creation
and destruction of holes and doublons. In recent years the similar models 
have been studied intensively~\cite{2a} - \cite{12}. In particular, some of
these models~\cite{2a,2b,4,4a,5} have been solved exactly under the 
condition that the number of doubly occupied sites is conserved.

The important puzzle arising in an investigation of the generalized models is 
metal-insulator transition problem.
In papers~\cite{2,13} a new mean-field approximation (MFA) 
which leads to the correct description of metal-insulator transition (MIT)
has been proposed. 
In the present paper we use this MFA to study MIT in a generalized Hubbard 
model with correlated hopping at half-filling and zero temperature. 
In Sec.~2 we
introduce the Hamiltonian of narrow-band model. The decoupling scheme of 
Green
functions is described in Sec.~3. Also single-particle Green function 
and energy spectrum of electon system are calculated. In Sec.~4 the 
expression for energy gap width is found. With the help of this formula and
the obtained expression for the concentration of polar states MIT is studied.
Finally, Sec.~5 is devoted to the conclusions.

\section{Hamiltonian of narrow-band model}
\setcounter{equation}{0}
Theoretical analysis, on the one hand, and 
available experimental data, on the other hand, point out the necessity of 
the Hubbard model~\cite{14} generalization by taking into account 
interelectron interactions describing intersite hoppings of electrons 
(correlated hopping)~\cite{1,14a,14b}. The characteristic property of these
correlated hopping integrals is the dependence on the occupation of sites by
electrons. 

So, we start from the following generalization of the Hubbard model
including correlated hopping~\cite{1,2}:
\begin{eqnarray}
H=&&-\mu \sum_{i\sigma}a_{i\sigma}^{+}a_{i\sigma}+
{\sum \limits_{ij\sigma}}'a_{i\sigma}^{+}\left(t_0+\sum_k J(ikjk)n_k\right)
a_{j\sigma}
\nonumber\\
&&+U \sum_{i}n_{i\uparrow}n_{i\downarrow},
\end{eqnarray}
where $a_{i\sigma}^{+}, a_{i\sigma}$ are the operators of creation and 
destruction for the electrons with spin $\sigma$ ($\sigma=\uparrow, 
\downarrow$) on $i$-site, $n_{i\sigma}=a_{i\sigma}^{+}a_{i\sigma}$, 
$n_i=n_{i\uparrow}+n_{i\downarrow}$; $\mu$ is the chemical potential; 
$U$ is the intra-atomic Coulomb repulsion;
$t_0$ is the matrix element which describe the hoppings of electrons
between nearest-neighbor sites of lattice in consequence of electron-ion 
interaction, 
\begin{eqnarray}
J(ikjk)=\int \int \phi^*({\bf r}-{\bf R}_i)\phi({\bf r}-{\bf R}_j)
{e^2\over |{\bf r}-{\bf r'}|}|\phi({\bf r'}-{\bf R}_k)|^2{\bf drdr'},
\end{eqnarray}
($\phi$-function is the Wannier function). 
The prime at second sum in Eq.~(2.1) signifies that $i\neq{j}$. 

In Hamiltonian~(2.1) we rewrite the sum 
$\sum'_{ij\sigma k}J(ikjk)a^{+}_{i\sigma}n_ka_{j\sigma}$
in the form
\begin{eqnarray}
{\sum_{ij\sigma}}'\sum_{\stackrel{k\neq{i}}{k\neq{j}}}J(ikjk)a^{+}_{i\sigma}
n_ka_{j\sigma}+{\sum_{ij\sigma}}'\left(J(iiij)a^{+}_{i\sigma}a_{j\sigma}
n_{i{\bar \sigma}}+h.c.\right)
\end{eqnarray}
(${\bar \sigma}$ denotes the spin projection which is opposite to $\sigma$),
here we have used that $J(iiji)=J(jiii)=J(iiij)$ in consequence of the 
matrix elements symmetry. Let us suppose (as in the papers~\cite{1,2}) that
\begin{eqnarray}
{\sum_{ij\sigma}}'\sum_{\stackrel{k\neq{i}}{k\neq{j}}}J(ikjk)a^{+}_{i\sigma}
n_ka_{j\sigma}=T_1{\sum_{ij\sigma}}'a^{+}_{i\sigma}a_{j\sigma}
\end{eqnarray}
with $T_1=n\sum_{\stackrel{k\neq{i}}{k\neq{j}}}J(ikjk)$
and $n=\langle n_{i\uparrow}+n_{i\downarrow}\rangle$ (sites $i$ and $j$ are
nearest neighbors); it should be noted that this supposition is exact in
the homeopolar limit ($n_i=1$).

Thus at half-filling ($n=1$) we can write Hamiltonian~(2.1) in the form
\begin{eqnarray}
H=&&-\mu \sum_{i\sigma}a_{i\sigma}^{+}a_{i\sigma}+
t{\sum \limits_{ij\sigma}}'a_{i\sigma}^{+}a_{j\sigma}+
T_2{\sum \limits_{ij\sigma}}' \left(a_{i\sigma}^{+}a_{j\sigma}n_{i\bar \sigma}+h.c.\right)
\nonumber\\
&&+U \sum_{i}n_{i\uparrow}n_{i\downarrow},
\end{eqnarray}
where $t=t_0+T_1$ and $T_2=J(iiij)$. 

Note that at $n\neq 1$ the model will be described also by Hamiltonian~(2.5)
with $t=t_0+nT_1$ i.e. taking into account the correlated hopping $T_1$ leads
to the concentration dependence of the hopping integral $t$. It is the
distinction of the present model from similar 
models~\cite{2a,2b,4,4a,14a,14b}. In the case $n=1$ 
taking into account the correlated hopping $T_1$ leads to the renormalization 
of the hopping integral $t_0$ only.

Rewrite Hamiltonian~(2.5) in terms of $X_i^{kl}$-Hubbard operators~\cite{15} 
using the formulae~\cite{16}
\begin{eqnarray*}
a_{i\uparrow}^{+}=X_i^{\uparrow 0}-X_i^{2\downarrow}, \ \
a_{i\uparrow}=X_i^{0\uparrow}-X_i^{\downarrow 2},
\\
a_{i\downarrow}^{+}=X_i^{2\uparrow}+X_i^{\downarrow 0}, \ \
a_{i\downarrow}=X_i^{\uparrow2}+X_i^{0\downarrow},
\end{eqnarray*}
where $X_i^{kl}$ is the transition-operator of $i$-site from state 
$|l\rangle$ to 
state $|k\rangle$; $|0\rangle$ denotes the site, which is not occupied
by an electron (hole), $|\sigma \rangle\equiv a_{i\sigma}^{+}|0\rangle$ 
denotes the singly occupied (by an electron with spin $\sigma$) $i$-site, 
$|2\rangle\equiv a_{i\uparrow}^{+}a_{i\downarrow}^{+}|0\rangle$ denotes the
doubly occupied (by two electrons with  the opposite spins) $i$-site (doublon).

In terms of $X_i^{kl}$-operators Hamiltonian~(2.5) takes the following
form:
\begin{eqnarray}
H=H_0+H_1+H'_1,
\end{eqnarray}
with
\begin{eqnarray*}
&&H_0=-\mu \sum_i \left(X_i^{\uparrow}+X_i^{\downarrow}+2X_i^2\right)+
U\sum_{i}X_i^2,\\
&&H_1=t{\sum \limits_{ij\sigma}}'X_i^{\sigma 0} X_j^{0\sigma} +
\tilde{t}{\sum \limits_{ij\sigma}}'X_i^{2\sigma}X_j^{\sigma 2},
\\
&&H'_1=t'{\sum \limits_{ij\sigma}}' \left(\eta_{\sigma}X_i^{\sigma 0}
X_j^{\bar{\sigma} 2}+h.c.\right),
\end{eqnarray*}
where $X_i^k=X_i^{kl}X_i^{lk}$ is the operator of the number of 
$|k\rangle$-states on $i$-site, $\eta_{\uparrow}=-1,\ \eta_{\downarrow}=1$;
\begin{eqnarray}
\tilde{t}=t+2T_2,\quad
t'=t+T_2.
\end{eqnarray}

The single-particle Green function
\begin{eqnarray}
G_{pp'}^{\sigma}(E)=\langle\langle a_{p\sigma}\!|\hfill a_{p'\sigma}^{+}
\rangle\rangle
\end{eqnarray}
in terms of Hubbard operators is written
\begin{eqnarray}
G_{pp'}^{\sigma}(E)=&&\langle\langle
X_p^{0\sigma}|X_{p'}^{\sigma 0}\rangle\rangle +\eta_{\sigma}\langle\langle
X_p^{0\sigma}|X_{p'}^{2\bar{\sigma}}\rangle\rangle +\eta_{\sigma}\langle\langle 
X_p^{\bar{\sigma} 2}|X_{p'}^{\sigma 0}\rangle\rangle \nonumber\\
&&+\langle\langle 
X_p^{\bar{\sigma} 2}|X_{p'}^{2\bar{\sigma}}\rangle\rangle.
\end{eqnarray}

$H_0$ describes the atomic limit of narrow-band models.

$H_1$ describes the translational hopping of holes and doublons. In the present
model (in contrast to the narrow-band models of the Hubbard type) the 
hopping integrals of holes $t$ and doublons $\tilde{t}$ are different.
It should be noted that in consequence of the difference of the hopping 
integrals, which describe translational hopping of current carriers within 
the lower (hole) band and upper (doublon) band, the energy width of the upper 
band can be much 
smaller and the effective mass of current carriers within this band can be 
much larger than in the lower band. 
Thus, within the proposed model the ideas of the ``wide'' and ``narrow'' 
subbands and the ``light'' and ``heavy'' current 
carriers are introduced (as the result of electron-electron interactions).

$H'_1$ describes the processes of paired creation and destruction of holes
and doublons.

\section{Energy spectrum of electron system: MFA}
\setcounter{equation}{0}

Here we use MFA proposed in papers~\cite{2,13}  to calculate 
the energy spectrum of electron system described by Hamiltonian~(2.7).

The Green function $\langle\langle X_p^{0\sigma}|X_{p'}^{\sigma 0}\rangle\rangle$
is given by the equation
\begin{eqnarray}
(E+\mu)\langle\langle X_p^{0\sigma}|X_{p'}^{\sigma 0}\rangle\rangle=&&
{\delta_{pp'}\over 2\pi}\langle X_p^{\sigma}+X_p^{0}\rangle+ 
\langle\langle\left[X_p^{0\sigma}, H_1\right]|X_{p'}^{\sigma 0}\rangle\rangle
\nonumber\\
&&+\langle\langle\left[X_p^{0\sigma}, H'_1\right]|X_{p'}^{\sigma 0}\rangle
\rangle,
\end{eqnarray} 
with $[A, B]=AB-BA$,
\begin{eqnarray}
\left[X_p^{0\sigma}, H_1\right]=t\sum_j\left((X_p^{\sigma}+X_p^{0})
X_j^{0\sigma}+X_p^{\bar{\sigma} \sigma}X_j^{0\bar{\sigma}}\right)-\tilde{t}
\sum_jX_p^{02}X_j^{2\sigma},
\end{eqnarray}
\begin{eqnarray}
\left[X_p^{0\sigma}, H'_1\right]=&&-t'\sum_jX_p^{02}X_j^{\bar{\sigma}0}+
t'\sum_jX_p^{\bar{\sigma} \sigma}X_j^{\sigma2}\nonumber\\
&&-t'\sum_j(X_p^{\sigma}+X_p^{0})X_j^{\bar{\sigma}2}.
\end{eqnarray}
To break off the sequence of Green function equations according to generalized
Hartree-Fock approximation~\cite{17} we suppose that
\begin{eqnarray}
\left[X_p^{0\sigma}, H_1\right]=\sum_{j}\epsilon(pj)X_j^{0\sigma},\
\left[X_p^{0\sigma}, H'_1\right]=\sum_{j}\epsilon_1(pj)X_j^{\bar{\sigma}2},
\end{eqnarray}
where $\epsilon(pj)$ and $\epsilon_1(pj)$ are the non-operator expressions.
The representation choice of the commutators in form~(3.2) and (3.3) is
prompted by the operator structure of these commutators which maps the energy
non-equivalence of the hopping processes prescribed by $H_1$ and $H'_1$.
Taking into account~(3.4) we rewrite Eq.~(3.1) in the form
\begin{eqnarray}
(E+\mu)\langle\langle X_p^{0\sigma}|X_{p'}^{\sigma 0}\rangle\rangle=&&
{\delta_{pp'}\over 2\pi}\langle X_p^{\sigma}+X_p^{0}\rangle+
\sum_{j}\epsilon(pj)\langle\langle X_j^{0\sigma}|X_{p'}^{\sigma 0}\rangle
\rangle
\nonumber\\
&&+\sum_{j}\epsilon_1(pj)\langle\langle X_j^{\bar{\sigma}2}|
X_{p'}^{\sigma 0}\rangle\rangle.
\end{eqnarray}
After anticommutation of both sides of the first of formulae~(3.4) with
$X_{k}^{\sigma 0}$ and the second formula with $X_k^{2\bar{\sigma}}$ we obtain
\begin{eqnarray}
\epsilon(pk)(X_{k}^{\sigma}+X_{k}^{0})=&&t(X_{p}^{\sigma}+X_{p}^{0})
(X_{k}^{\sigma}+X_{k}^{0})+tX_{k}^{\sigma\bar{\sigma}}
X_{p}^{\bar{\sigma}\sigma}
\nonumber\\
&&-\delta_{pk}t\sum_{j}X_{k}^{\bar{\sigma}0}X_{j}^{0\bar{\sigma}}
+\delta_{pk}\tilde{t}\sum_{j}X_{j}^{2\sigma}X_{k}^{\sigma 2}
\nonumber\\
&&-\tilde{t}X_{k}^{2 0}X_{p}^{0 2},
\end{eqnarray}
\begin{eqnarray}
\epsilon_1(pk)(X_{k}^{\bar{\sigma}}+X_{k}^{2})=&&-t'(X_{p}^{\sigma}+X_{p}^{0})
(X_{k}^{\bar{\sigma}}+X_{k}^{2})+t'X_{p}^{\bar{\sigma}\sigma}
X_{k}^{\sigma\bar{\sigma}}
\nonumber\\
&&-\delta_{pk}t'\sum_{j}X_{j}^{\bar{\sigma}0}X_{k}^{0\bar{\sigma}}
+\delta_{pk}t'\sum_{j}X_{k}^{2\sigma}X_{j}^{\sigma 2}
\nonumber\\
&&-t'X_{k}^{2 0}X_{p}^{0 2}.
\end{eqnarray}

Similarly, for the Green function $\langle\langle X_{p}^{\bar{\sigma}2}| 
X_{p'}^{\sigma 0}\rangle\rangle$ we can write the equation
\begin{eqnarray}
(E+\mu-U)\langle\langle X_{p}^{\bar{\sigma}2}| X_{p'}^{\sigma 0}\rangle
\rangle=&&\sum_{j}\tilde{\epsilon}(pj)\langle\langle X_{j}^{\bar{\sigma}2}| 
X_{p'}^{\sigma 0}\rangle\rangle\nonumber\\
&&+\sum_{j}\epsilon_2(pj)\langle\langle
X_{j}^{0\sigma}| X_{p'}^{\sigma 0}\rangle\rangle,
\end{eqnarray}
where $\tilde{\epsilon}(pj)$ and $\epsilon_2(pj)$ are determined
through the expressions which are analogous to (3.6) and (3.7). 
Thus we obtain the closed system of equations for the Green functions 
$\langle\langle X_{p}^{0\sigma}| X_{p'}^{\sigma 0}\rangle\rangle$ and
$\langle\langle X_{p}^{\bar{\sigma}2}| X_{p'}^{\sigma 0}\rangle\rangle$. 

By neglecting correlated hopping and by averaging expressions~(3.6) and~(3.7) 
we obtain the approximations~\cite{14,18,19}; the defects of 
these approximations are 
well-known (see, for example Ref.~\cite{20}). Here we use the approach 
which has been proposed in the papers~\cite{2,13}.

To determine $\epsilon(pj),\ \epsilon_1 (pj)$ we rewrite $X_i^{kl}$-operator 
in Eqs.~(3.6) and~(3.7) in the form~\cite{20a} $X_i^{kl}=\alpha_{ik}^{+}\alpha_{il}$,
where $\alpha_{ik}^{+},\ \alpha_{il}$ are the operators of creation and 
destruction for $|k\rangle$- and $|l\rangle$-states
on $i$-site respectively (the Schubin-Wonsowsky operators~\cite{21}); 
thus $X_i^0=\alpha^{+}_{i0}\alpha_{i0},\ 
X_i^2=\alpha^{+}_{i2}\alpha_{i2},\ 
X_i^{\sigma}=\alpha^{+}_{i\sigma}\alpha_{i\sigma}$.
Let us substitute $\alpha$-operators by $c$-numbers in Eqs.~(3.6) and (3.7)
(here there is a partial equivalence with slave boson method~\cite{21a})
\begin{eqnarray}
\alpha_{i\sigma}^{+}=\alpha_{i\sigma}=\left({1-2d\over 2}\right)^{1/2},
\qquad
\alpha_{i 0}^{+}=\alpha_{i 0}=\alpha_{i 2}^{+}=\alpha_{i 2}=d^{1/2}
\end{eqnarray}
(we consider a paramagnetic case, electron concentration on site $n=1$); 
$d$ is the concentration of polar states (holes or doublons). 

The proposed approximation is based on
the following physical idea. Let us consider a paramagnetic Mott-Hubbard
insulator at $T\neq 0$. Within the wide temperature interval ($k_BT\ll U$)
the concentration of polar states is small ($d\ll 1$). 
An analogous consideration is valid for a paramagnetic Mott-Hubbard semimetal 
(hole and doublon subbands overlap weakly, $d\ll 1$).
So, the change of 
states and polar excitations influences on $|\sigma\rangle$-states weakly. 
Thus we may consider $|\sigma\rangle$-states as the
quasiclassical system and substitute the operators $\alpha^{+}_{i\sigma},\
\alpha_{i\sigma}$ by $c$-numbers. In addition, when we find 
$\epsilon(pj),\ \epsilon_1 (pj)$ we substitute the creation and destruction 
operators of $|0\rangle$- and $|2\rangle$-states through the respective 
quasiclassical 
expressions. Actually the proposed approximation is
equivalent to a separation of the charge and spin degrees of freedom. Note
that the present approach is the most justifiable when $d\to 0$.  

Thus in $\bf{k}$-representation we obtain
\begin{eqnarray}
\epsilon({\bf k})=(1-2d+2d^2)t_{\bf k}-2d^2\tilde{t}_{\bf k}, \quad
\epsilon_1({\bf k})=-2dt'_{\bf k},
\end{eqnarray}
where $t_{\bf k},\ \tilde{t}_{\bf k},\ t'_{\bf k}$ are the Fourier 
transforms of the hopping integral $t,\ \tilde{t},\ t'$ respectively.
Similarly, we find that
\begin{eqnarray}
\tilde{\epsilon}({\bf k})=(1-2d+2d^2)\tilde{t}_{\bf k}-2d^2t_{\bf k}, \quad
\epsilon_2({\bf k})=-2dt'_{\bf k}.
\end{eqnarray}

The Fourier transform of the Green function 
$\langle\langle X_{p}^{0\sigma}| X_{p'}^{\sigma 0}\rangle\rangle$
is found from the system of equations~(3.5) and~(3.8)
\begin{eqnarray}
\langle\langle X_{p}^{0\sigma}| X_{p'}^{\sigma 0}\rangle\rangle_{\bf k}=
{1\over 4\pi}\cdot {E+\mu-U-(1-2d+2d^2)\tilde{t}_{\bf k}+2d^2t_{\bf k}\over 
(E-E_1({\bf k}))(E-E_2({\bf k}))},
\end{eqnarray}
with
\begin{eqnarray}
&&E_{1,2}({\bf k})=-\mu+{(1-2d)(t_{\bf k}+\tilde{t}_{\bf k})+U\over 2}\mp 
{1\over 2}F_{\bf k},
\\
&&F_{\bf k}=\sqrt{\left[B(t_{\bf k}-\tilde{t}_{\bf k})-U\right]^2+
(4dt'_{\bf k})^2},\
B=1-2d+4d^2.
\end{eqnarray}

An analogous procedure is realized also in the equations for the other Green 
functions in Eq.~(2.6).

Finally, in {\bf k}-representation the single-particle Green function is
\begin{eqnarray}
&&G_{\bf k}(E)={1\over 2\pi}\left({A_{\bf k}\over E-
E_1({\bf k})}+{B_{\bf k}\over E-E_2(\bf k)}\right),\\
&&A_{\bf k}={1\over 2}-{2dt'_{\bf k}\over F_{\bf k}}, \quad
B_{\bf k}={1\over 2}+{2dt'_{\bf k}\over F_{\bf k}}.
\end{eqnarray} 

Single-particle Green function~(3.15) gives the exact atomic and
band limits: if $U=0$ and $t_{\bf k}=\tilde{t}_{\bf k}=t'_{\bf k}=
t_0({\bf k})$ (it means neglecting correlated hopping) then 
$G_{\bf k}(E)$ takes the band form ($d=1/4$ when $U=0$), 
if $t_{\bf k}=\tilde{t}_{\bf k}=t'_{\bf k}\rightarrow 0$ then we obtain the 
exact atomic limit.

The peculiarities of obtained quasiparticle energy spectrum~(3.13) of 
narrow-band system 
which is described by Hamiltonian~(2.5) are the dependence on the 
concetration of
polar states and the non-equivalence of the lower and upper Hubbard bands. 
This non-equivalence is caused by the difference of the hopping integrals $t$, 
$\tilde{t}$, $t'$.

Quasiparticle energy spectrum~(3.13) allows to study MIT in the proposed
model.

\section{Metal-insulator transition}
\setcounter{equation}{0}

With the help of energy spectrum of electrons~(3.13) we find the expression
for the energy gap width (difference of energies between bottom of the upper 
and top of the lower Hubbard bands):
\begin{eqnarray}
&&\Delta E=-(1-2d)(w+\tilde{w})+{1\over 2}(Q_1+Q_2),
\\
&&Q_1=\sqrt{\left[ B(w-\tilde{w})-U\right]^2+(4dzt')^2},\nonumber\\ 
&&Q_2=\sqrt{\left[ B(w-\tilde{w})+U\right]^2+(4dzt')^2},\nonumber 
\end{eqnarray}
where $w$ and $\tilde{w}$ are the half-widths of the lower (hole) and upper 
(doublon) Hubbard bands respictevely: $w=z|t|,\ \tilde{w}=z|\tilde{t}|$ 
($z$ is the 
number of nearest neighbors to a site).

The peculiarities of the expression for energy gap~(4.1) are dependences on
the concentration of polar states, on the widths of hole and doublon bands, 
on the hopping integral $t'$ (thus on external pressure). At given $U,\ t,\ 
\tilde{t},\ t'$ (constant external pressure) the concentration dependence of
$\Delta E$ allows to study MIT under the action of external
influences: temperature change, photoeffect and magnetic field. In particular,
$\Delta E(T)$-dependence can lead to the transition from a metallic state to
an insulating state with the increase of temperature~\cite{21b} 
(in this connection the transition from the state of a paramagnetic metal to 
the paramagnetic insulator state in the (V$_{1-x}$Cr$_x$)$_2$O$_3$ 
compound~\cite{22,23}, in NiS$_2$~\cite{24} and in 
the NiS$_{2-x}$Se$_x$ system~\cite{24,24a,24b} should be noted).
Under the action of light or magnetic field the concentration of polar 
states can be changed; it leads to
the fact that the energy gap width is changed also and MIT can occur.

Distinction of formulae~(3.13)--(3.16), (4.1) from earlier obtained results
(e.g., see reviews~\cite{23,25,26}) is the dependence on concentration of 
polar states. Let us find the expression for its calculation.

The concentration of polar states is given by the equation
\begin{eqnarray}
d=\langle X_i^2\rangle =&&{1\over N}{\sum_{\bf k}\int\limits_{-\infty}
^{+\infty}}J_{\bf k}(E)dE
\nonumber\\
&&={1\over 2N}\sum_{\bf k}\left(
{C_{\bf k}\over \exp{E_1({\bf k})\over \theta}+1}+
{D_{\bf k}\over \exp{E_2({\bf k})\over \theta}+1}\right),
\end{eqnarray} 
where 
\begin{eqnarray*}
&&C_{\bf k}={1\over 2}-{B(\tilde{t}_{\bf k}-t_{\bf k})\over 2F_{\bf k}}-
{U\over 2F_{\bf k}}, \\
&&D_{\bf k}={1\over 2}+{B(\tilde{t}_{\bf k}-t_{\bf k})\over 2F_{\bf k}}+
{U\over 2F_{\bf k}},
\end{eqnarray*}
$\theta=k_BT$, $k_B$ is the Boltzmann's constant, $N$ is the number of sites, 
$J_{\bf k}(E)$ is the spectral intensity of the Green function
\begin{eqnarray}
\langle\langle X_p^{{\bar \sigma}2}|X_{p'}^{2{\bar \sigma}}\rangle
\rangle_{\bf k}={1\over 4\pi}\left({C_{\bf k}\over E-E_1({\bf k})}+
{D_{\bf k}\over E-E_2(\bf k)}\right).
\end{eqnarray}

At $T=0$ and the rectangular density of states
\begin{eqnarray*}
{1\over N}\sum_{\bf k}\delta(E-t_{\bf k})={1\over 2w}\theta(w^2-E^2)
\end{eqnarray*}
($\theta(x)=1$ if $x>0,\ =0$ otherwise) from Eq.~(4.2) we obtain that
\begin{eqnarray}
-{B\over z}{\tilde{t}-t\over \lambda}\left[\varphi (\epsilon_0)-\varphi 
(-\epsilon_0) \right]+{U\over z\sqrt{\lambda}}\left(1-{B^2
(\tilde{t}-t)^2\over \lambda}\right)\times\hfill
\nonumber\\
\times \ln\left|{\sqrt{\lambda}\varphi (\epsilon_0)- 
\lambda \epsilon_0 -BU(\tilde{t}-t)
\over \sqrt{\lambda}\varphi (-\epsilon_0)+\lambda \epsilon_0 -
BU(\tilde{t}-t)}\right|=8d-2 \quad (U< w+\tilde{w})
\end{eqnarray}
with
\begin{eqnarray*}
&&\epsilon_0=2\sqrt{\mu U-\mu^2\over (1-2d)^2(t+\tilde{t})^2-\lambda},\quad 
\mu={(1-2d+2d^2)w-2d^2\tilde{w}\over (1-2d)(w+\tilde{w})}U, 
\nonumber\\
&&\varphi(\epsilon)=\left\{ \lambda \epsilon^2-2BU(\tilde{t}-t)\epsilon+
U^2\right\}^{1\over 2}, \quad 
\lambda =B^2(\tilde{t}-t)^2+(4dt')^2.
\end{eqnarray*}
For narrow-band semimetal ($d\ll 1$) Eq.~(4.4) takes the following form:
\begin{eqnarray}
d={1\over 4}\left({1-{U\over w+\tilde{w}}}\right).
\end{eqnarray}

Fig.~1 shows the dependence of $d$ on $U/w$ which is obtained from 
Eq.~(4.4). The parameters $\tau_1=T_1/|t_0|,\ \tau_2=T_2/|t_0|$ characterize
the value of correlated hopping. One can see that a value of $d$ 
depends on the parameters of correlated hopping $\tau_1,\ \tau_2$ 
(thus on $\tilde{w}/w$) weakly when $U/w$ is close to zero. But with the 
increase of $U/w$ the 
concentration of polar states becomes strongly dependent on the parameters $\tau_1,\ 
\tau_2$. It testifies on the fact that taking into account
the correlated hopping is important to consider the metal-insulator transition 
problem.

Fig.~1 shows also that if $U\geq w+\tilde{w}$ then the concentration of polar 
states $d=0$. In the special case $t+\tilde{t}=t'=0$ this consequence is in
accordance with the results of Refs.~\cite{4,4a,8}.
 
At $T=0$ the energy gap width $\Delta E\leq 0$ (i.e. MIT occurs) 
when the condition
\begin{eqnarray}
U\leq w+\tilde{w}
\end{eqnarray}
is satisfied (in agreement with general physical ideas~\cite{23}). For the
special case $t'=0$ condition~(4.6) covers the exact results of 
Refs.~\cite{2b,4,5}. 

Fig.~2 which is obtained from formula~(4.1) using  Eq.~(4.4) shows
that in a metallic state the overlapping of energy subbands decreases and
in an insulating state the energy gap width increases with decrease of 
the parameter $\tilde{w}/w$ (at given $U/w$).

In the Hubbard model energy gap width~(4.1) takes the following form:
\begin{eqnarray}
\Delta E=-2w(1-2d)+\sqrt{U^2+(4dw)^2},
\end{eqnarray}
and the concentration of polar states~(4.4) is
\begin{eqnarray}
d=\left({1\over 4}+{U\over 32dw}\ln(1-4d)\right)\theta(2w-U).
\end{eqnarray}
In the region of metal-insulator transition $d=1/4-U/(8w)$; this dependence 
is in qualitative accordance with the result of Brinkman and Rice~\cite{26a}
obtained by use of Gutzwiller variational method~\cite{26b}, those of
the general Gutzwiller-correlated wave functions in infinite 
dimensions~\cite{27} and the Kotliar-Ruckenstein slave bosons~\cite{21a}. 
For $U/2w\to 0$ 
we obtain $d=1/4+U/(8w)\ln(U/2w)$ (if we consider Coulomb repulsion as
perturbation then $d(U\to 0)=1/4-{\cal O}(U)$); in order to compare the 
obtained 
dependence~(4.8) $d$ on $U/w$ in the Hubbard model with other approximate 
theories see e.g.~\cite{28}). $\Delta E\leq 0$ when the condition 
$2w\geq U$ is satisfied. 

\section{Conclusions}

In the present paper Mott-Hubbard transition has been studied in a generalized 
Hubbard model with correlated hopping at half-filling using a mean-field 
approximation~\cite{2,13}. 

We have obtained the expression to calculate the concentration of polar states.
With the increase of the 
parameter
$\tilde{w}/w$ ($0\leq \tilde{w}/w\leq 1$) the concentration of polar states 
increases at given $U/w$.  

Quasiparticle energy spectrum has been calculated. With the help of this 
energy spectrum we have found the energy gap width. The peculiarities of the
obtained expressions are dependence on the concentration of polar states and
the non-equivalence of the upper and lower Hubbard bands. The increase of the 
parameter
$\tilde{w}/w$ (at given $U/w$) leads to decreasing energy gap width. 
In consequence of it, in particular, MIT occurs when
the condition $U/w=1+\tilde{w}/w$ is satisfied.

The cases $n\neq1$ and $T\neq0$, an application of the obtained results 
to the interpretation of the experimental data will be considered in the next 
papers.  

The authors are grateful to Prof. I.~V.~Stasyuk for valuable discussions.

Figure 1: Concentration of polar states $d$ as a function of $U/w$: the 
upper curve corresponds to $\tau_1=\tau_2=0$; the middle 
curve -- $\tau_1=\tau_2=0.1$; 
the lower curve -- $\tau_1=\tau_2=0.2$.

Figure 2: Energy gap width $\Delta E$ as a function of $U/w$: the upper 
curve corresponds to $\tau_1=\tau_2=0.2$; the lower curve -- 
$\tau_1=\tau_2=0$.

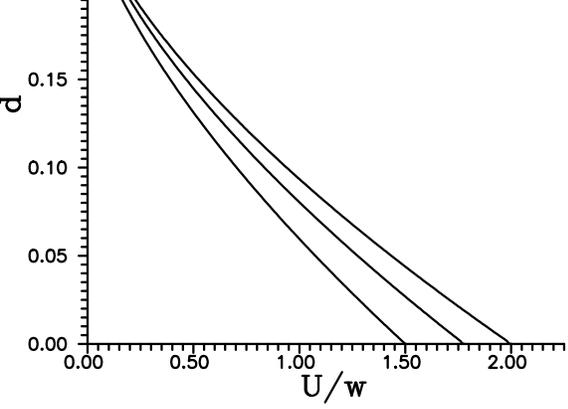
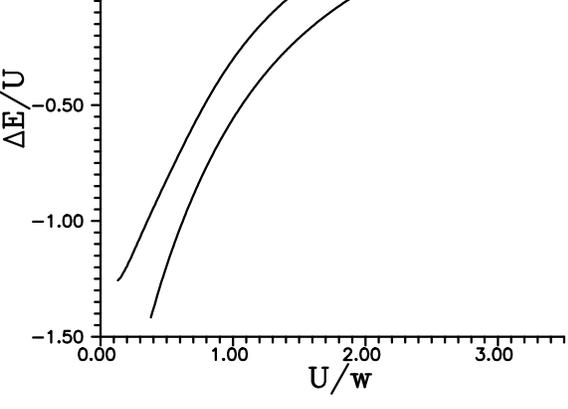
\end{document}